# Path Conditions Help to Locate and Localize Faults from Programs

Safeeullah Soomro[1], Zahid Hussain[2], Ayaz Keerio[3]
[1]Department of Computer Science and Engineering,
[1]Yanbu University College, Yanbu Al-Sinaiyah, Kingdom of Saudi Arabia
Email: safeeullah.soomro@yuc.edu.sa,

[2]Quaid-e-Awam University of Engineering, Science and Technology, Pakistan
Email: zhussain@quest.edu.pk

[3]University of Sindh, Jamshoro, Pakistan
Email: ayazkeerio@hotmail.com

**Abstract.** Precisely and automatically detection of faults in programs, is a software engineering dream. Every effort in this regard takes us one step closer to realizing it. Many efforts have been taken from the people of these areas on testing, verification and debugging. We are proposing such effort for the research community of this domain is using path conditions to generate a minimal set of PLOFC (possible lines of faulty code). It's a runtime method that will effectively bring the minimal possible set of faulty lines of code through the help of path conditions and some heuristics involved. In this paper we are generating possible fault locations from programs using path conditions which can put positive impact on the static analysis of programs. Further we discuss the basic ideas regarding path conditions, the theory, and first analysis results. This work is based on a previous work that uses the variable dependences for fault detection. We showed some examples to the applicable of this idea and can be useful for software verification.

## 1  Introduction

Software verification is the process to cheek the program or the software if work in correct way. The verification can prove a program does not have an error to make all properties workable in a useful way. Some formal tools are to be used to make error free programs which are using carefully. This approach is generally used to start from the beginning to development process. The verification has two kinds of process, which are assertion and invariant. During the verification process, one or more persons are involved to find errors and second are used to verify these root causes. Also second part can helps the user to prevention the failure of the project. The Verification part comes before Validation, which incorporates Software inspections, reviews, audits, walkthroughs, buddy checks etc. There are three levels of software verification: Software Unit Testing, Software Integration Testing and Software System Testing. There are four types of verification that can be applied to levels: Inspection, Analysis, Testing and Demonstration. As an example, the most effective way to find anomalies at the component level is inspection. We will discuss about formal inspections process.

Software formal inspections are in-process technical reviews of a product of the software life cycle conducted for the purpose of finding and eliminating defects. Formal inspections were developed by Michael Fagan at IBM to improve software quality and increase programmer productivity. They are conducted by individuals from development, test, and assurance, and may include users.

Formal verification has been introduced various different techniques. In general formal verification is the act of proving or disproving the correctness of a system with respect to a certain formal properties. The author [1] defines Model checking with respect to formal verification for every program state in the light of program specifications. The authors [2, 3, 4] define formal verification as proving properties in a system. In this technique we have programs specifications to find faults in program. These specifications compare with program states with different states and finds inconsistency between them. The Model checking [1] pinpoints the faults between all states. According to formal methods, formal verification is defined as final assessment of system which are concerning with properties. Proving each specification of system is a very hard to prove it, so major drawback of formal verification is the lack of formal specification in more cases. Writing post-conditions as program specification[6] is counted hard as writing correct computer program. So it takes too much time and money on these formal specifications[5], mainly software's are developing in these days without them.

In this paper we are presenting the method to locate and localize the faults from programs using path conditions [9]. These path conditions are presented by Snelting [9] for the program analysis. We [7, 8] extended this work towards locating the faults using some examples. We are presenting graphs to show that how we can use of these conditions to follow the real misbehavior's comes through value output. In this paper section 2 presents the examples of diagnosis to prove our idea to locating the faults from procedural programs. Section 2 shows that how we are localizing faults in programs. After that section 4 represents the related research and finally we conclude the paper and specify the future research.

## 2     Diagnosis Example

In this section we are presenting the example which shows that the faults can be diagnose through our path conditions. Path conditions are using to evaluate all kinds of statement towards the end of program which were presented by Snelting[9]. We are using these conditions for locating faults from programs. Basically we are making sections from programs and make graphs to find the variables which cannot be affect the results so slice them from the programs. Other variables are compulsory to watch for analysis. After that we are asserting the values as model checkers works to make assumptions to find the exact value of output for the verification of any code. After finding the variables who can affect the final outcomes so that we pinpoint and diagnosis [6] these values behind variables lines. So we found the real lines which can affect the output and presented as real misbehavior of programs. It's a runtime method that will effectively bring the minimal possible set of faulty lines of code through the help of path conditions [9] and some heuristics involved [7,8]. These are the steps which can create blocks of statements and driving diagnosis with the help of path conditions and applying diagnosis theory to detect faults from programs.

1. Creating blocks from execution of statements.
2. Minimizing blocks for fault detection and fault localization.
3. Driving diagnosis with the help of Path Conditions to acquire a minimal PLOFC.
4. Analysis of data obtained from step 3

**Figure 1: Example #1**

```
1     x1 = a;
2     y1 = b;
3     if (x1 < y1)
4           then z1 = x1 + 2
5           else  z1 = y1 + 2
6     z1 = z1 + y1;
7     if (y1 > 5)
8           then z1 = z1 + 5
9           else z1  = z1 - 2
10    z1 = z1 + 3;
```
Input Values of variables are a=3 and b=4

Tracing example number 1 for checking each and every value of variables using path execution number 1 that are as under: - execution of path number 1 is executed. Then the values of example number 1 are as under:-

$x1 = 3$
$y1 = 4$
$z1 = 3 + 2 = 5 + 4 = 9 + 5 = 14 + 3 = 17.$

The specific value of $z1 = 17.$ Suppose the value of line number 4 is in the example number 1 that is

$z1 = x1 + 4.$

So the output will be changed and we don't get specified output. Here is the tracing program number 1 due to changed value

  x1 = 3
  y1 = 4
  z1 = 3 + 4 = 7 + 4 = 11 + 5 = 16 + 3 = 19

Now output of z1 = 19 and it is changed from specified output so we can trace the each and every statement using path conditions to diagnose the fault from program.

## Step 1: Creating Blocks from execution of statements

Having obtained a path we can create blocks following method creates blocks in a manner that is most suitable for further diagnosis. For Path (1, 10) =? Blocks can be

**Blocks:**

There can be 7 blocks for example #1
  A, B, –B, C, D, –D and E
  Block A contains lines:       1, 2
  Block B contains lines:       3, 4
  Block -B contains lines:      3, 5 (notice line number 3 replicated)
  Block C contains lines:       6
  Block D contains lines:       7, 8
  Block -D contains lines:      7, 9 (notice line number 7 replicated)
  Block E contains lines: 10

**All Path Formula:**

A ^ (B v -B) ^ C ^ (D v -D) ^ E
All path formula in CNF (conjunctive normal form) satisfies program execution.

**All Possible Execution Paths obtained:**

|     | In Blocks | Lines in Blocks |
| --- | --- | --- |
| (1) | A ^ B ^ C ^ D ^ E | 12 ^ 34 ^ 6 ^ 78 ^ 10 |
| (2) | A ^ B ^ C ^ -D ^ E | 12 ^ 34 ^ 6 ^ 79 ^ 10 |
| (3) | A ^ -B ^ C ^ D ^ E | 12 ^ 35 ^ 6 ^ 78 ^ 10 |
| (4) | A ^ -B ^ C ^ -D ^ E | 12 ^ 35 ^ 6 ^ 79 ^ 10 |

**All Path Execution Flow Graph:**

**Graph 1: Path Execution Flow**

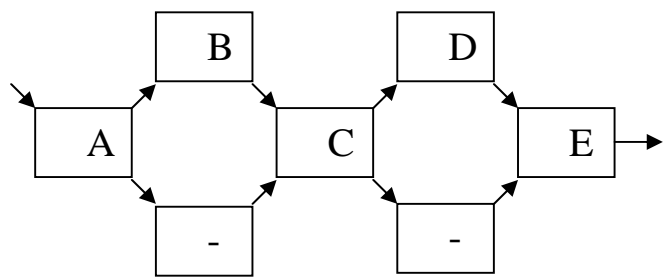

## Step 2: Minimizing blocks for fault detection and fault localization.

Note that our target variable for diagnosis is z1. And our method uses path conditions to detect and localize fault at runtime. For minimizing blocks we give some definitions as follows:-

**Definition 1: Blocks whose output variable is not our target variable and none of the variables dependent on our target variables can be safely removed.**

Sadly in this example no such blocks can be removed, because either all block contain the target variable z1 or variables dependent on target variable z1 i.e. x1 and y1. With my blocking scheme and large programs I am determined that several blocks can be removed with this definition applied.

**Definition 2: Blocks that are not in the execution path can be safely removed**

Blocks highlighted with green color are part of execution path number 1 and contain possible faulty lines of code; all other blocks marked with red are removed from our diagnosis.

**Graph 2: Runtime Path Execution Flow**

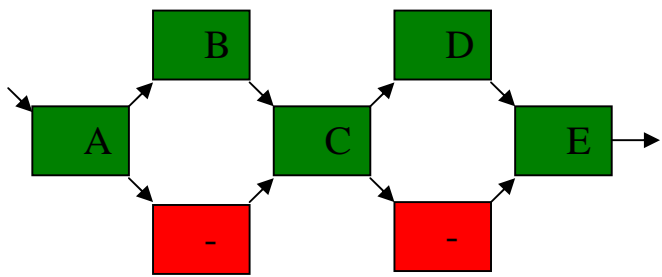

In both sections we have proved that how path conditions are important to make analysis of programs. Also give hint to add previous block statements and ignore un-wanted blocks. Hence Path conditions are useful for program debugging and verification.

## Step 3: Driving diagnosis with the help of Path Conditions

From path conditions we have obtained a reduced form of example 1 looks like following

**Figure 2: Example #1 (reduced)**

1      x1 = a;

```
2       y1 = b;
3       if (x1 < y1)
4           then z1 = x1 + 2
6       z1 = z1 + y1;
7       if (y1 > 5)
8           then z1 = z1 + 5
10      z1 = z1 + 3;
```
Notice line number 5 and 9 have been removed using path condition diagnosis by definitions provided in Step1.

For the remaining part we create data dependencies for each line

```
1       x1 = a;                     (x1 on a)
2       y1 = b;                     (y1 on b)
3       if (x1 < y1)
4           then z1 = x1 + 2        (z1 on x1) (z1 on 2) (z1 on y1)
6       z1 = z1 + y1;               (z1 on z1) (z1 on y1)
7       if (y1 > 5)
8           then z1 = z1 + 5  (z1 on z1) (z1 on) (z1 on 5) (z1 on 5)
10      z1 = z1 + 3;                (z1 on z1) (z on 3)
```

By using the composition operator we obtain a set
         **{(x1, a), (y1, b), (z1, x1), (z1, y1), (z1, z1)}** --------------------------------- 1

**Definition 1: Give definitions to undefined constants in the program, associate unique names to constants thus clearly identifying them.**

Example number 1 has four constants need to be assigned unique variables.
  $c_1 = 2$
  $c_2 = 5$ (line number 7)
  $c_3 = 5$ (line number 8)
  $c_4 = 3$

**Figure 3: Example #1 (unique naming of constants)**

```
c1 = 2
c2 = 5
c3 = 5
c4 = 3
1    x1 = a ;                    (x1 on a)
2    y1 = b ;                    (y1 on b)
3    if (x1 < y1 )
4        then z1 = x1 + c1     (z1 on x1) (z1 on c1) (z1 on y1)
6    z1 = z1 + y1 ;              (z1 on z1) (z1 on y1)*
7    i f (y1 > c2)
8        then z1 = z1 + c3     (z1 on z1)* (z1 on y1)* (z1 on c2) (z1 on c3)
10   z1 = z1 + c4 ;              (z1 on z1)* (z on c4)
```

Figure 3 shows the code after unique naming of constants and updating the dependencies accordingly.
By using the composition operator on new dependencies we obtain a set
 { (x1,a), (y1,b), (z1,x1), (z1,c1), (z1,y1), (z1,z1), (z1,c3), (z1,c4) } --------------- 2
**Definition 2: Associate dependencies with operators. Leaving = operator on same variable i.e. x = x not required, also leaving control operators because they don't have a role to play after execution path is available.**

With the above definition the example number 1 becomes

**Figure 4: Example #1 (operator association with dependencies)**

```
c1 = 2
c2 = 5
c3 = 5
c4 = 3
1      x1 = a ;                        (x1 on a with =)
2      y1 = b ;                        (y1 on b with =)
3      i f (x1 < y1 )
4      then
       z1 = x1 + c1      (z1 on x1 with =) (z1 on c1 with +)
                         (z1 on x1 with <)* (z1 on y1 with <)*
6      z1 = z1 + y1 ;    (z1 on z1 with =)* (z1 on y1 with +)
7      i f (y1 > c2)
8      then z1 = z1 + c3     (z1 on z1 with =)* (z1 on c3 with +)
                              (z1 on y1 with >)* (z1 on c2 with >)*
10     z1 = z1 + c4 ;                  (z1 on z1 with =)* (z on c4 with +)
```

Note: dependences marked with * are not allowed in the set by definition. Our Final set becomes

{ (=,(x1,a)), (=,(y1,b)), (=,(z1,x1)), (+,(z1,c1)), (+, (z1,y1)), (+,(z1,c3)), (+,(z1,c4)) }
By mapping this new set over path execution flow graph we obtain

**Graph 3: Path Execution Flow**

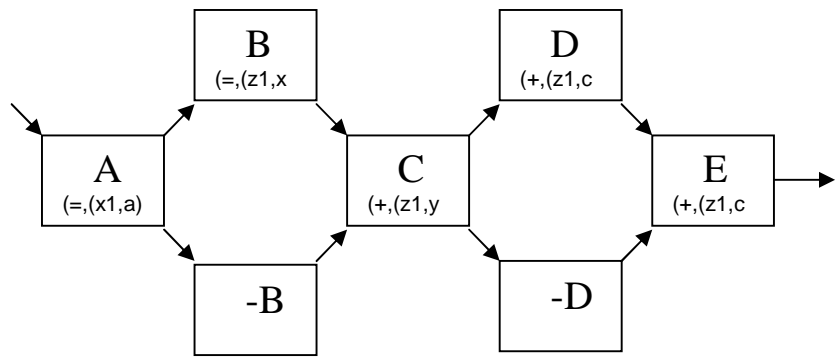

For our example we used input a, b and we got result $z_1 = 17$ and we assumed correct result to be $z_1 = 19$. So now we can make a function Predict Faulty Lines that will run some algorithms and show possible faulty lines.

**Predict-Faulty-Lines (Inputs, Output, Desired-Output, Path-Condition-Graph)**

**Step 4: Analysis of data obtained from step 3**

The data obtained from path execution flow graph can be used for two types of analysis
1. For reducing dependencies or further localizing faults
2. For making/suggesting corrections in faulty code

**Example 1: An Algorithm for reducing dependencies**

In this particular scenario from graph 3; the input a is directly being assigned to $x_1$ and $y_1$ and since then they have not been updated/changed by either any local variable or any local constant. This logical argument intuitively suggests such dependencies to be removed from faulty set of dependencies[5, 6]. So we can safely remove the dependency $(=, (x_1, a)), (=, (y_1, b)), (=, (z_1, x_1)), (+, (z_1, y_1))$, saying that an error in either of these dependencies will notify an error with input but we assumed the input to be correct.
The algorithm reduces dependencies into **{(+, ($z_1$, $c_1$)), (+, ($z_1$, $c_3$)), (+, ($z_1$, $c_4$))} -- 3**

**Example 2: An algorithm for making/suggesting corrections in faulty code**

We take the two outputs (desired and target) and we take their difference with mod i.e. $|z_1' - z_1| = |17 - 19| = |-2| = 2$ (here $z_1'$ is desired output). So output difference is OD = 2. There are three constants in the final dependencies set and no local variables. So now what we can do is with each constant add and subtract the OD and verify the output as follows

$\{c_1 + 2, c_1 - 2, c_3 + 2, c_3 - 2, c_4 + 2, c_4 - 2\}$

We will perform this analysis for the number of dependencies of +, - operator associations. We have to run this test 2d times where d is the number of dependencies. For the above changes three of the test result into a match in the outputs

$\{c_1 - 2, c_3 - 2, c_4 - 2\}$

This particular algorithm is verifying and suggesting corrections as well.

## Conclusions

In this paper we have shown the importance path conditions, how they can be used effectively, how to use it with data dependencies. Also we have introduced another concept of operator association with data dependencies [7, 8, 10] and shown their importance for different analysis algorithms. For the flow graph (graph 3) that I produced, intuitively and accurately predicts the faulty blocks, these blocks can be reduced by analyzing their output variables. Using abstract dependencies alone is not sufficient enough for runtime debugging and the use of operator associations seems the obvious intuitive solution available. We have shown using small example drive diagnosis and how to use it to localize faults.

Future work has to implement GUI to make tool for locating and localizing faults from programs using Java.

## References

1. de Kleer, J., Williams, B. C.: A pragmatic survey of automatic debugging., Springer LNCS 749, 1--15 (1993)
2. Edmung, O.G, Clarke, M. A. Doron.: Frequency –Model Checking MIT Press, Cambrige, Massachussets (1999)
3. Hinchy, M.B, Bowen, J.P.: Applied for formal methods, Prentice Hall, London (1995).
4. Jeanette,M. W.: Specifier's introduction to formal methods. Computers, Volume. 8. pp, 29-10-22 Charlston, South Carolina, Sept (1990)
5. Stumptner, M., Wotawa, F.: Java Diagnosis Experiments -- Status and Outlook; IJCAI '99 Workshop on Qualitative and Model Based Reasoning for Complex Systems and their Control Stockholm, Sweden (1999)
6. Reiter, R.: A theory of Diagnosis from first principles. Artificial Intelligence 32(1), 57--95 (1987)
7. Soomro, S.: Using abstract dependences to localize faults from procedural programs. Proceedings of AIA, pp. 180--185, Innsbruck, Austria (2007)
8. Soomro, S., Wotawa, F.: Detect and Localize Faults in Alias-free Programs using Specification Knowledge. Proceedings of IEAAIE, LNCS, Springer (2009)
9. Snelting, G, Robschink, T, Krinke, J.: Efficient Path Conditions in Dependence Graphs for Software Safety Analysis, ACM Transactions on Software Engineering and Methodology, **Vol. 15**, (4), pp. 410--457, October (2006).
10. Wotawa, F., Soomro, S.: Using abstract dependencies in debugging. Proceedings of Qualitative Reasoning QR-05 (2005).